\documentclass[fleqn,usenatbib]{mnras}

\usepackage{newtxtext,newtxmath}
\usepackage[normalem]{ulem}

\usepackage[T1]{fontenc}
\usepackage{ae,aecompl}
\usepackage[normalem]{ulem}

\usepackage{graphicx}	
\usepackage{amssymb}	
\usepackage{hyperref}
\hypersetup{
  colorlinks   = true, 
  urlcolor     = blue, 
  linkcolor    = blue, 
  citecolor   = blue 
}
\usepackage{xcolor}

\newcommand\ion[2]{#1$ ${\scshape{#2}}} 
\newcommand{\kms}{\,km\,s$^{-1}$}	    
\newcommand{\caps}[1]{{\scshape{#1}}}  
\def\msun{M$_{\odot}$}
\def\pnv{V1838 Aql}

\def\pnv{V1838~Aql}
\def\apj{ApJ}
\def\aj{AJ}
\def\mnras{MNRAS}
\def\pasp{PASP}
\def\aap{A\&A}
\def\apjs{ApJS}
\def\msun{M$_{\odot}$}
\def\kms{km s$^{-1}$}
\def\porb{$P_{orb}$}

\title[The bow-shock nebula in V1838 Aql]{From outburst to quiescence: spectroscopic evolution of V1838 Aql imbedded in a bow-shock nebula}

\author[Hern\'andez Santisteban et al.]{J. V. Hern\'andez Santisteban$^{1,2}$\thanks{e--mail: 
	j.v.hernandez@uva.nl}, J. Echevarr\'{\i}a$^{2,1}$, S. Zharikov$^{3}$, V.  Neustroev$^{4}$, 
 \newauthor G. Tovmassian$^{3}$, V. Chavushyan$^{5}$, R. Napiwotzki$^{6}$, R.Costero$^{2}$,  R. Michel$^{3}$,  
\newauthor L. J. S\'anchez$^{2}$, A. Ruelas--Mayorga$^{2}$, L. Olgu\'in$^{7}$, Ma. T. Garc\'ia--D\'iaz$^{3}$, 
\newauthor D. Gonz\'alez--Buitrago$^{3,8}$, E. de Miguel$^{9}$, E. de la Fuente$^{10}$,  R. de Anda$^{2}$,
\newauthor and V. Suleimanov$^{11,12,13}$\\
$^{1}$Anton Pannekoek Institute for Astronomy, University of Amsterdam, Science Park 904, NL--1098 XH Amsterdam, the Netherlands\\
$^{2}$ Instituto de Astronom\'ia, Universidad Nacional Aut\'onoma de M\'exico, Apartado Postal 70--264, \\
~~Ciudad Universitaria, M\'exico D.F., C.P. 04510, Mexico\\
$^{3}$ Instituto de Astronom\'{\i}a, Universidad Nacional Aut\'onoma de M\'exico, Ensenada, Baja California, C.P. 22830, Mexico\\
$^{4}$ Astronomy Research Unit, University of Oulu, PO Box 3000, FIN-90014, Finland \\
$^{5}$ Instituto Nacional de Astrof\'isica, \'Optica y Electr\'onica, Apartado Postal 51, CP 72000, Puebla, Mexico \\
$^{6}$ Centre for Astrophysics Research, Science and Technology Research Institute, University of Hertfordshire, Hatfield AL10 9AB, UK\\
$^{7}$ Departamento de Investigaci\'on en F\'{\i}sica, Universidad de Sonora, Blvd. Rosales y Colosio, 83190 Hermosillo, Sonora, Mexico \\
$^{8}$ Department of Physics and Astronomy, University of California, Irvine, California 92697, USA \\
$^{9}$ Departamento de Ciencias Integradas, Universidad de Huelva, E--21071 Huelva, Spain\\
$^{10}$ Instituto de Astronom\'{\i}a y Meteorolog\'{\i}a, Departamento de F\'{\i}sica, CUCEI, Guadalajara, Jalisco, 44100, Mexico\\
$^{11}$Institut f\"ur Astronomie und Astrophysik, Universit\"at T\"ubingen, Sand 1, D-72076 T\"ubingen, Germany\\
$^{12}$Kazan (Volga region) Federal University, Kremlevskaya str. 18, 420008 Kazan, Russia\\
$^{13}$Space Research Institute of the Russian Academy of Sciences, Profsoyuznaya Str. 84/32, Moscow 117997, Russia
}

\date{Accepted XXX. Received YYY; in original form ZZZ}

\pubyear{2019}

\begin{document}

\date{Submitted: \today, ArXiv: \today}

\pagerange{\pageref{firstpage}--\pageref{lastpage}} \pubyear{2018}

\maketitle

\label{firstpage}

\begin{abstract}

We analyse new optical spectroscopic, direct-image and X-ray observations of the recently discovered a high proper motion cataclysmic variable \pnv. 
The data were obtained during its 2013 superoutburst and its subsequent quiescent state. 
An extended emission around the source was observed up to 30 days after the peak of the superoutburst, interpreted it as a bow--shock formed by a quasi-continuous outflow from the source in quiescence. The head of the bow--shock is coincident with the high--proper motion vector of the source ($v_{\perp}=123\pm5$ \kms) at a distance of $d=202\pm7$ pc. 
The object was detected as a weak X-ray source ($0.015\pm0.002$ counts s$^{-1}$) in the plateau of the superoutburst, and its flux lowered by two times in quiescence (0.007$\pm$0.002 counts s$^{-1}$).
Spectroscopic observations in quiescence we confirmed the orbital period value $P_{\rm{orb}}=0.0545\pm 0.0026$ days, consistent with early-superhump estimates, and the following orbital parameters: $\gamma= -21\pm3$ km s$^{-1}$ and $K_1 = 53\pm3$ km s$^{-1}$. 
The white dwarf is revealed as the system approaches quiescence, which enables us to infer the effective temperature of the primary $T_{eff}=11,600\pm400$K. The donor temperature is estimated $\lesssim 2200$K and suggestive of a system approaching the period minimum. 
Doppler maps in quiescence show the presence of the hot spot in \ion{He}{I} line at the expected accretion disc-stream shock position  and an unusual structure of the accretion disc in H$\alpha$. 

\end{abstract}



 
\section{Introduction}
\label{intro}

Cataclysmic Variables (CVs) are close binary systems where a white dwarf (WD) accretes from a low--mass star via Roche--lobe overflow, often creating an accretion disc \citep[for a review see][]{war95}. The evolution of CVs is driven by the removal of angular momentum from the system, which leads to the depletion of the donor and causes the orbital period ($P_{orb}$) to shrink. This process continues until the donor reaches the sub-stellar regime (i.e. a brown dwarf donor) \citep{how97}, where its internal structure causes the donor to expand in response to the loss mass, thus leading to an increase in \porb. This fact causes a sharp cut-off in the \porb\ distribution called the period minimum \citep{pas81}. In addition, the long time-scales associated at the period minimum leads to the accumulation of most CVs between 1-2 hrs \citep{gan09}, with a large fraction of systems evolving towards longer \porb, known as period bouncers. Out of this short \porb\ population, up to ${\sim}70$ \%  \citep{kab99,gol15} should be period bouncers and harbour a sub-stellar donor. Theoretically, the stellar to sub-stellar transition roughly coincides with the period minimum. However, there is little observational evidence of its location given the lack of detailed characterisation of systems around the period minimum \citep[e.g.][]{lea06,har16,her16,neu17,pala18}. 

Short orbital period systems (often referred as WZ Sge--type objects) are characterised by enhanced brightness, extended outburst duration (${\sim}30$ days) as well as the onset of superhumps (low-amplitude variability close the orbital period of the system) shortly after maximum brightness. These are often classified as superoutbursts, to distinguish them from those observed in classical dwarf novae. Therefore, the follow-up and detailed characterisation of these systems is paramount to confirm the nature of the donor and test against theoretical expectations \citep{kni11}. 

The discovery of a new transient, \pnv ~(also known as PNV~J19150199+0719471), was initially reported by Koichi Itagaki on May 31 2013, as a possible Nova reaching V$\sim$10 mag, who reported also that the object was below 15.5 mag on his unfiltered survey image taken on 21.608 UT\footnote{\url{www.cbat.eps.harvard.edu/unconf/followups/J19150199+0719471.html}}. 
Kato (vsnet--alert 15776) pointed out that the high proper motion made it a good candidate for a WZ Sge--type nearby star, close to the galactic plane, ($\ell = 42.145^{\circ}$, $b = -1.865^{\circ}$). This classification was confirmed in subsequent vsnet--alerts, where various stages of superhumps were observed as the superoutburst evolved, and a mean superhump period of 0.05803(1) days was reported  by Kato (vsnet--alert 15931). 

\begin{figure*}
\includegraphics[trim=0.2cm 0.5cm .2cm 0.0cm,clip,width=17cm]{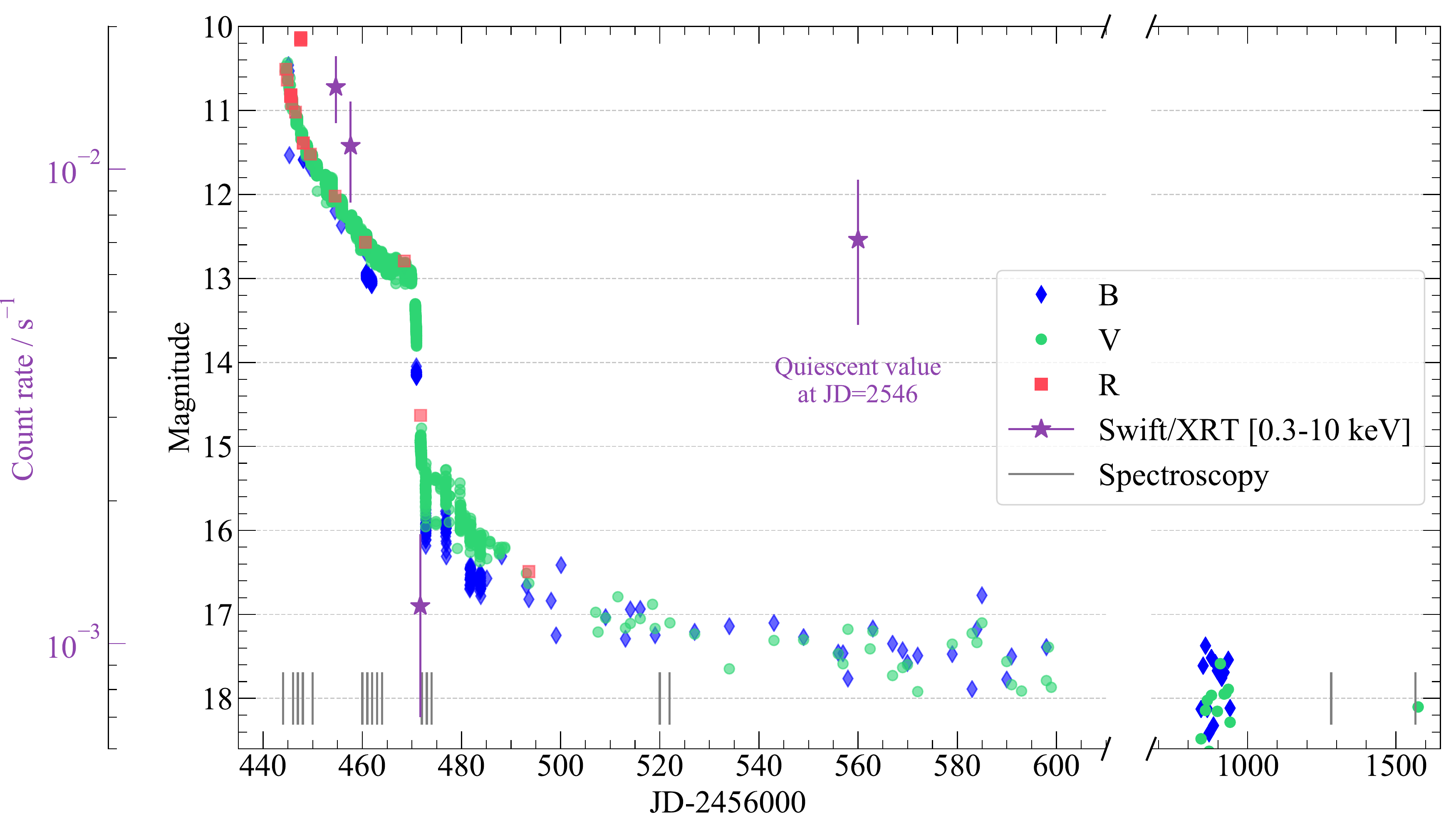}
\caption{
AAVSO and \textit{Swift}/XRT light curves during and after the 2013 superoutburst of \pnv\ . The observations were taken in V (green dots), R (red squares) and B (blue diamonds) standard filters. The grey vertical marks indicate the epochs of our spectroscopic observations. The 0.3-10 keV \textit{Swift}/XRT count rate is shown in purple stars. Note that the magnitude of the star is still above the minimum derived from the pre-outburst magnitude V=18.6 at the time of the GTC spectral observations on day 1566. The last \textit{Swift/XRT} point is shown as a reference
}
\label{fig:aavso-figs}
\end{figure*}

In this paper we present and discuss CCD direct images, X-ray observations and extensive low and high dispersion spectroscopy during outburst and quiescence of \pnv. We show a most striking result: the detection of a nebulosity in emission during the course of the superoutburst. In addition, we present time-resolved spectroscopy and determine orbital parameters from the emission lines, as well as Doppler  tomography. The presence of a white dwarf revealed at quiescence allowed us to determine the white dwarf (WD) temperature. Finally, we discuss whether the system is approaching or receding from the minimum orbital period.

\begin{table*}
  \setlength{\tabcolsep}{0.1em}
  \begin{center}
    \caption{Log of photometric and spectroscopic observations}
    \label{tab:speclog}
    \begin{tabular}{lccccc}
       \hline \hline
       \noalign{\smallskip}
Spectroscopy & Julian Date & Range~~~ & No. of & Exposure & Comments\\
Date & (2456000+) & (\AA)~~~~~ & Spectra & Time (s) &\\
        \noalign{\smallskip}
       \hline
       \noalign{\smallskip}
2013 June 3     & 444     & 3900--7300 & 14 & 300 & Echelle \\
2013 June  5--7 & 446--448 & 4130--7575 & 29 & 300 & Boller \& Chivens \\
2013 June 9     & 450     & 5510--6730 & 29 & 300 & Boller \& Chivens \\
2013 June 16--19& 460--464 & 3900--7300 & 17 & 900 & Echelle \\
2013 June 28    & 472     & 4500--5700 & 14 & 420 & Boller \& Chivens \\
2013 June 29    & 473     & 5500--6700 & 14 & 420 & Boller \& Chivens \\
2013 June 30    & 474     & 4500--5700 & 14 & 420 & Boller \& Chivens  \\
2013 August 15  & 520     & 4000--7000 & 10 & 900 & Boller \& Chivens \\
2015 September 17 & 1282  & 4000--7500 & 03 & 1800& Boller \& Chivens  \\
2016 June 27    & 1566    & 3440--4610 & 01 & 600 & Osiris -- GTC \\
2016 June 27    & 1566    & 4500--6000& 02 & 245& Osiris -- GTC \\
2016 June 27    & 1566    & 5575--7685& 02 & 254& Osiris -- GTC \\
2016 June 27    & 1566    & 7330--10000& 01 & 600& Osiris -- GTC \\
2016 June 27    & 1566    & 5575--7685& 30 & 235& Osiris -- GTC \\
       \hline
       \hline

Imaging & Julian Date & Filter & FWHM & Exposure & Comments\\
Date & (2456000+) &  & (\AA) & Time (s) &\\
\hline
2013 June 3    & 444  & V & 980 & 30 \\
2013 June 18   & 462  & H$\alpha$ & 11 &  1260 & 180 s $\times$ 7 images \\
2013 June 29   & 473 & H$\alpha$ & 11 &  1800&  180 s $\times $ 10 images \\
2013 June 29   & 473 & Continuum (6650\AA) & 70  &  1800  &  180 s $\times $ 10 images\\
2013 June 29   & 473 & [OIII] 5007   & 52 &  1800 &  180 s $\times $ 10 images \\
2013 June 29   & 473 & [NII] 6583   & 10 &  1800 &  180 s $\times $ 10 images \\ \hline
Multicolor Band Photometry &  & Filters & Telescope & Site & Comments\\
& &  &  &  &\\
\hline
2016 November 9 & 1701 & BVRI    & NOT    & ORM       & Section\\
2017 March 8    & 1820 & JHK$_s$ & NOT    & ORM       & \ref{sec:imag}\\
2017 October 6  & 2032 & BVRiz   & NTT    & La Silla  & for more\\
2017 October 6  & 2032 & JHK$_s$ & NTT    & La Silla  & details\\
\hline

       \noalign{\smallskip}
 \end{tabular}
  \end{center}
\label{tab:log}
\end{table*}

\section{Observations and Reduction}
\label{sec:observations}
We will present the multi-wavelength observations of \pnv\ from its discovery to its quiescent state, spanning over $\sim4$ years. In order to simplify the reference to specific dates or epochs throughout the superoutburst, we make use of truncated Julian Days of the form HJD -- 2456000.
\subsection{Spectroscopy}
Observations were obtained with the Echelle spectrograph attached to the 2.1m Telescope of the Observatorio Astron\'omico Nacional at San Pedro M\'artir, on the nights of 2013 June 3 and June 16--19. The Marconi--2, a $2048\,\times\,2048$ detector, was used to obtain a spectral resolution of R${\sim}19\,000$, All  observations were carried out with the 300~{\it l/mm} cross--dispersor, which has a blaze angle at around 5500~\AA. The spectral coverage was about $3900$--$7300 $ \AA. The exposure time for each spectrum was 300~s for June 3 and 900~s for June 16--19. Low dispersion spectroscopy was also obtained on 2013 June 5--7 and 9 with the Boller~\&~Chivens spectrograph (B\&Ch), with two different gratings. On the first three nights, a 400 {\it l/mm} grating was used, to obtain a high S/N ratio in order to study the spectral energy distribution of the system, with a broad wavelength coverage of about $4130$--$7575$ \AA.  The exposure time for each spectrum was 300~s. In addition (on June 9) a 1200 {\it l/mm} grating was used, to obtain a higher spectral resolution around the interval $5510$--$6730$ \AA. In this setup, the exposure time for each spectrum was 300~s. 

Further observations were obtained with the B\&Ch during the nights of 2013 June 28, 29 and 30, with the 1200~{\it l/mm} grating to cover the range around H$\alpha$ (June 29) and H$\beta$ (June 28 and 30).  Additional low resolution observations were obtained also with the B\&Ch on 2013 August 15 ($4000$--$7000$ \AA~coverage) with an exposure time of 900~s per spectrum, and two years later on 2015 September 17 ($4000$--$7500$ \AA~coverage) with an exposure time of 1800~s per spectrum, both with the 400 {\it l/mm} grating and when the system had already dropped to V$\sim$ 17 mag. All observations were made with a 1\farcs 5 slit oriented in the EW direction. Arc spectra were taken frequently for wavelength calibration. 

The optical spectroscopy of \pnv\ in quiescence was obtained on 2016 June 27 (HJD=2457566) using the long--slit mode of  the OSIRIS instrument attached to the GTC 10.4m  telescope. Single spectra were obtained first to cover the optical region $\lambda\lambda= 3400-10000$ \AA\  with the R2500U (600~s), R2500V (500~s), R2500R (500~s), and R2500I (600~s) volume--phased holographic gratings (see bottom spectra in Fig.~\ref{fig:spec_evolution}). Phase--resolved spectroscopy was then obtained with the R2005R grating, in the wavelength interval $5575-7685$ \AA, centred  around the H$\alpha$ emission line. The exposure time for each spectrum was 235 sec, with a total coverage of around one and a half orbital cycles. Standard data reduction procedures for all spectroscopic observations were performed using the \caps{iraf}\footnote {\caps{iraf} is distributed by the National Optical Observatories, operated by the Association of Universities for Research in Astronomy, Inc., under cooperative agreement with the National Science Foundation.} software. The log of all spectroscopic observations is shown in Table~\ref{tab:log}.

\subsection{Photometry and narrow-band imaging}
\label{sec:imag}
Direct deep narrow filters images in H$\alpha$, H$\alpha$ Continuum, [\ion{O}{iii}] 5007 \AA, and [\ion{N}{ii}] 6583 \AA\ were obtained during 2013, June 3, 18 and 29 at the 0.84m Telescope of the Observatorio Astron\'omico Nacional at San Pedro M\'artir, Mexico. The log of all imaging observations is also presented in Table~\ref{tab:log}.

We also performed multi-colour optical and near-infrared photometry.  On 2016 November 8, we obtained BVRI images with the Andalucia Faint Object Spectrograph and Camera (ALFOSC) at the 2.56~m Nordic Optical Telescope (NOT), in the Observatorio de Roque de los Muchachos (ORM, La Palma, Spain). The integration times were 300 s in each filter. We observed the target again on 2017 October 6 with the New Technology Telescope (NTT) at La Silla Observatory, Chile. The images were captured with the ESO Faint Object Spectrograph and Camera (EFOSC2 -- \citealt{BuzzoniEFOSC}) through the BVRiz filters with exposure times of 40, 40, 40, 60 and 120 sec, respectively. In addition, we obtained two sets of near-infrared (NIR) observations with the JHK$_s$ filters. The observations were performed on 2017 March 8 with the NOTcam instrument on the NOT, and on 2017 Oct 6 with SOFI on the NTT \citep{MoorwoodSOFI}.  The log of all photometric observations is also shown in Table~\ref{tab:speclog}. 

Pre-outburst observations were derived from the Pan-STARRS1 database\footnote{The Pan-STARRS1 Surveys (PS1) and the PS1 public science archive have been made possible through contributions by the Institute for Astronomy, the University of Hawaii, the Pan-STARRS Project Office, the Max-Planck Society and its participating institutes, the Max Planck Institute for Astronomy, Heidelberg and the Max Planck Institute for Extraterrestrial Physics, Garching, The Johns Hopkins University, Durham University, the University of Edinburgh, the Queen's University Belfast, the Harvard-Smithsonian Center for Astrophysics, the Las Cumbres Observatory Global Telescope Network Incorporated, the National Central University of Taiwan, the Space Telescope Science Institute, the National Aeronautics and Space Administration under Grant No. NNX08AR22G issued through the Planetary Science Division of the NASA Science Mission Directorate, the National Science Foundation Grant No. AST-1238877, the University of Maryland, Eotvos Lorand University (ELTE), the Los Alamos National Laboratory, and the Gordon and Betty Moore Foundation.}.
The mean epoch of the observations is around day 29 in our notation (i.e on 2012, April 12), more than a year earlier to the superoutburst.
We extracted the data using a small radius of 0.03 arcmin and obtained the following mean PSF magnitudes in the g, r, i, z, and y filters: 18.56 $\pm$ 0.04, 18.67 $\pm$ 0.01, 18.76 $\pm$ 0.03, 18.61 $\pm$ 0.05 and 18.43 $\pm$ 0.07, respectively. 

\subsection{X-rays}
\pnv\ was also observed five times with the Neil Gehrels Swift Observatory \citep{geh04}. These observations were taken in the middle of the superoutburst plateau stage on 2013 June 11 and 14 (the total exposure time of this data subset is 5.35 ks), at the end of the rapid fading stage on June 28 ($\sim$5.13 ks), and in quiescence on 2019 March 4 ($\sim$2.85 ks). During the plateau stage, \textit{Swift}-XRT detected a weak X-ray source with a count-rate of 0.0149$\pm$0.0024 counts s$^{-1}$, which is dropped to the level of about 0.0012$\pm$0.0006 counts s$^{-1}$ at the end of the rapid fading stage. In quiescence, however, the count-rate has been found at the level of 0.0071$\pm$0.0022 counts s$^{-1}$, that is lower than during the superoutburst plateau but is higher than during the rapid fading stage (see Table~\ref{tab:xraylog} for the observation log and Fig.~\ref{fig:aavso-figs}). This pattern --- the outburst flux is higher than in quiescence with a deep dip during the outburst rapid fading --- is in contrast to ordinary dwarf novae which usually show a depression of the X-ray flux during outbursts, but is in agreement with X-ray observations of WZ Sge-type stars \citep[see][and references therein]{neu17bb}. 

\begin{table}
  \setlength{\tabcolsep}{0.1em}
  \begin{center}
    \caption{Log of \textit{Swift}/XRT observations}
    \label{tab:xraylog}
    \begin{tabular}{rcccr}
       \hline \hline
       Julian Date &   Exp. Time  & Obs. ID & X-ray count rate\\
       2456000+ &   ks  &  & count s$^{-1}$\\
       \hline
454.658 &    3.370 &00032861001/2  & 0.0149$^{+0.0024}_{-0.0024}$\\
\rule{0pt}{4ex}
457.640 &    1.981 &00032861003  & 0.0112$^{+0.0027}_{-0.0027}$\\
\rule{0pt}{4ex}
471.689 &    5.127 &00032870001/15  & 0.0012$^{+0.0007}_{-0.0005}$\\
\rule{0pt}{4ex}
2546.656 &    2.853 &00032870016  & 0.0071$^{+0.0024}_{-0.0020}$\\
       \noalign{\smallskip}
      \hline
 \end{tabular}
  \end{center}
\label{tab:log_xray}
\end{table}

The plateau-, decline-stage, and quiescent spectra of \pnv\ consist of only 58, 6, and 12 counts, respectively, therefore no meaningful spectral analysis is possible. Nevertheless, assuming that the spectrum of \pnv\ is similar to other WZ~Sge-type stars such as SSS~J122221.7$-$311525 and GW~Lib \citep{neu17bb}, and using the count-rate as the scale-factor, one can estimate the X-ray flux of \pnv\ in the 0.3--10 keV energy range during the superoutburst plateau and in quiescence to be $\sim$4.5$\times$10$^{-13}$ and $\sim$2.3$\times$10$^{-13}$ erg\,s$^{-1}$cm$^{-2}$, respectively, with the corresponding luminosity of $\sim$2.3$\times$10$^{30}$ and $\sim$1.2$\times$10$^{30}$ erg\,s$^{-1}$ (adopting the distance of 202 pc, see Section~\ref{sec:bowshock}). 
These luminosities are in agreement with those found for other WZ~Sge-type stars and accreting white dwarfs \citep{Reis2013,neu17bb}.

\section{Discovery of a nebulosity during the early stages of outburst}
\label{sec:neb}

We report the detection of an extended component around the object shortly after maximum light. This nebulosity was first detected in the Echelle 
spectra on days 460--464, (see day 460 in Fig.~\ref{fig:bow_shock}, lower panel, two days before the first H$\alpha$ image). Further H$\alpha$ images were taken on day 462 which revealed a clear diffuse emission centred on the object, as seen in top two panels of Fig.~\ref{fig:bow_shock}. This explains the asymmetry of the line profile observed in the Echelle spectroscopy, due to the slit position (east--west) only covering a fraction of the extended structure at the  west of the object. We used \caps{daophot} \citep{ste87} to subtract the point sources, as shown in the middle panel of Fig.~\ref{fig:bow_shock}. Its morphology resembles that of a bow shock e.g. BZ Cam \citep{hea92}. Additional images were taken on day 473 with a narrow filters around the continuum near 6650~\AA ~and also with two narrow filters [OIII] 5007 \AA~and [NII] 6563 \AA. No emission from these forbidden lines was detected. The extended H$\alpha$ emission was present for nearly a month after the peak of the outburst with no apparent change in morphology. No nebular stage was detected later. 
 
\subsection{Proper motion and bow shock}
\label{sec:bowshock}

As mentioned in Section~\ref{intro}, \pnv\ shows a strong proper motion, initially observed from our field images obtained in 2013 and archival data from the Palomar survey (POSS--I and POSS--II). Recently, the {\it Gaia} DR2 release \citep{gaia16,gaia18} confirmed the high proper motion as well as provided a parallax for the source\footnote{\pnv\ catalogue ID is {\it Gaia} DR2 4306244746253355776}. 
The proper motion of \pnv\ is $\mu_{\alpha}\cos\delta=-90.1\pm0.2$ mas yr$^{-1}$ and $\mu_{\delta}= -91.4\pm0.2$ mas yr$^{-1}$. In conjunction with the parallax ($\pi=4.95\pm0.16$ mas), we performed a joint Bayesian inference for the distance and the tangential velocity following \citet{bailer18}\footnote{The R-based code is available at \url{https://github.com/ehalley/parallax-tutorial-2018}} which leads to a tangential velocity of $v_{\perp}= 123\pm5$ \kms\ and a distance of $202\pm7$ pc. The joint and marginal posterior distributions are shown in Appendix~\ref{sec:A1}. The distance inferred is consistent with our initial SED fitting estimates in quiescence presented in Section \ref{sec:wd}. Thus, combining the systemic radial velocity measurements (see Section \ref{sec:spec}) with the proper motion, we obtain a space velocity of $125\pm5$~km~s$^{-1}$. \pnv\ is the third CV with the highest transverse velocity in the Galaxy, just below SDSS 1507+22 ($v_{\perp}=182\pm30$ \kms), a confirmed metal--poor halo--binary \citep{pat08,uthas11}, and BF Eri, with an estimated value of $v_{\perp}\sim400$~\kms~ \citep{kle04,neu08}. Follow up H$\alpha$ observations are needed  to perform a more detailed study of its kinematics and origin, and in particular, far-- and near--ultraviolet observations are desired to search for anomalous line ratios that might indicate evidence for Population II membership.

\begin{figure}
\includegraphics[angle=0,trim=0 1.6cm 0 0,clip,width=0.96\columnwidth]{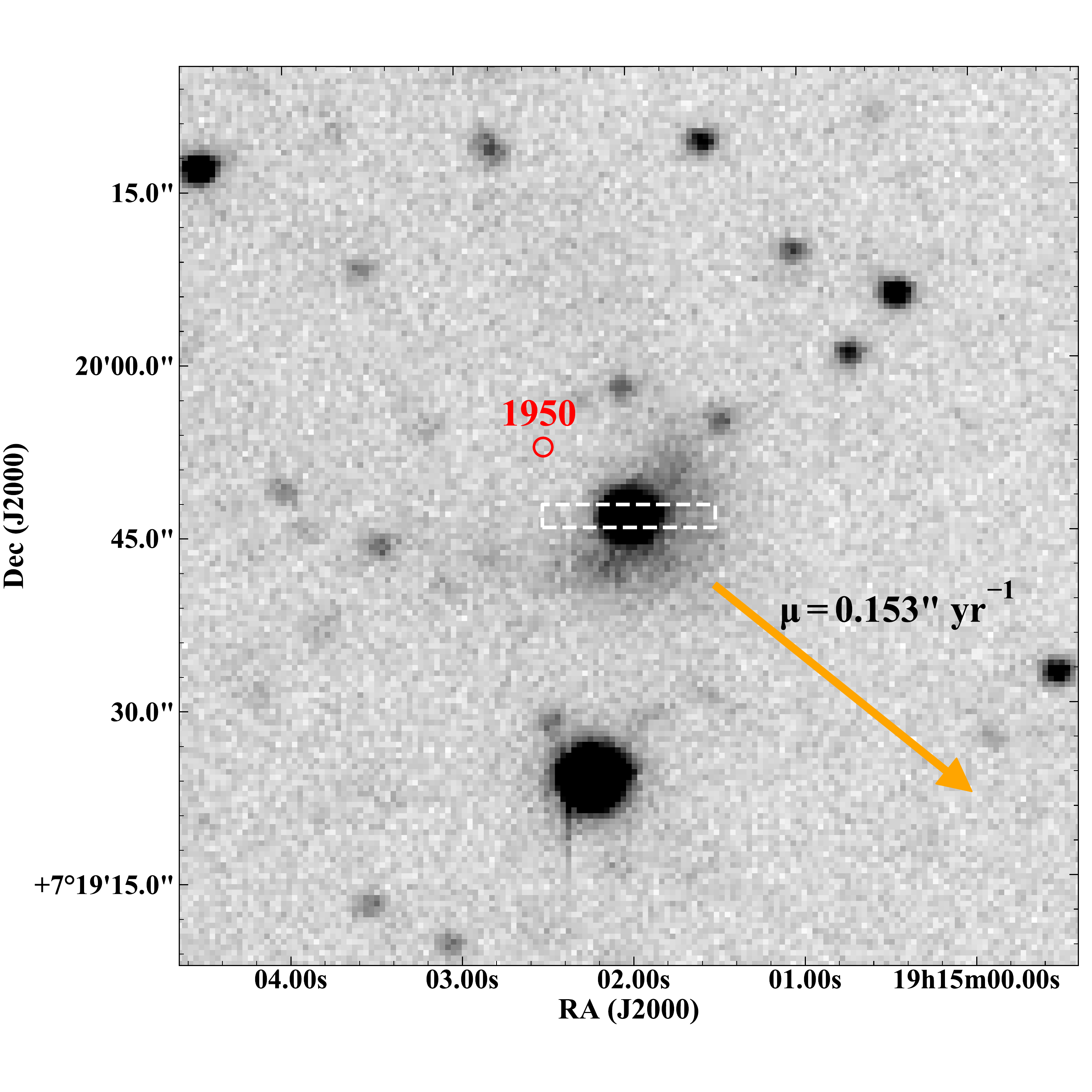}\\
\includegraphics[angle=0,trim=0 1.6cm 0 0,clip,width=0.96\columnwidth]{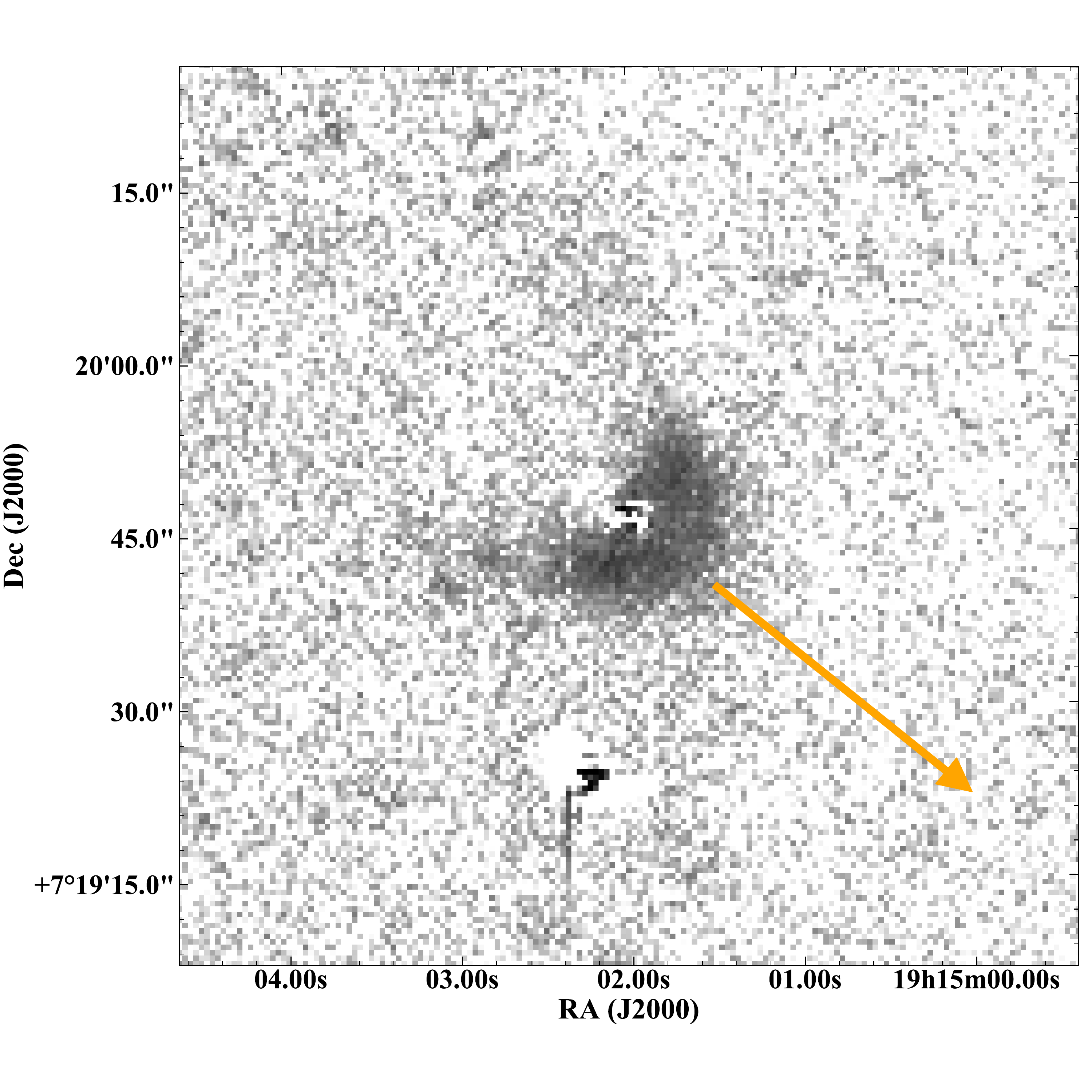}\\
\includegraphics[angle=0,trim=0 0.cm 0 0,clip,width=0.96\columnwidth]{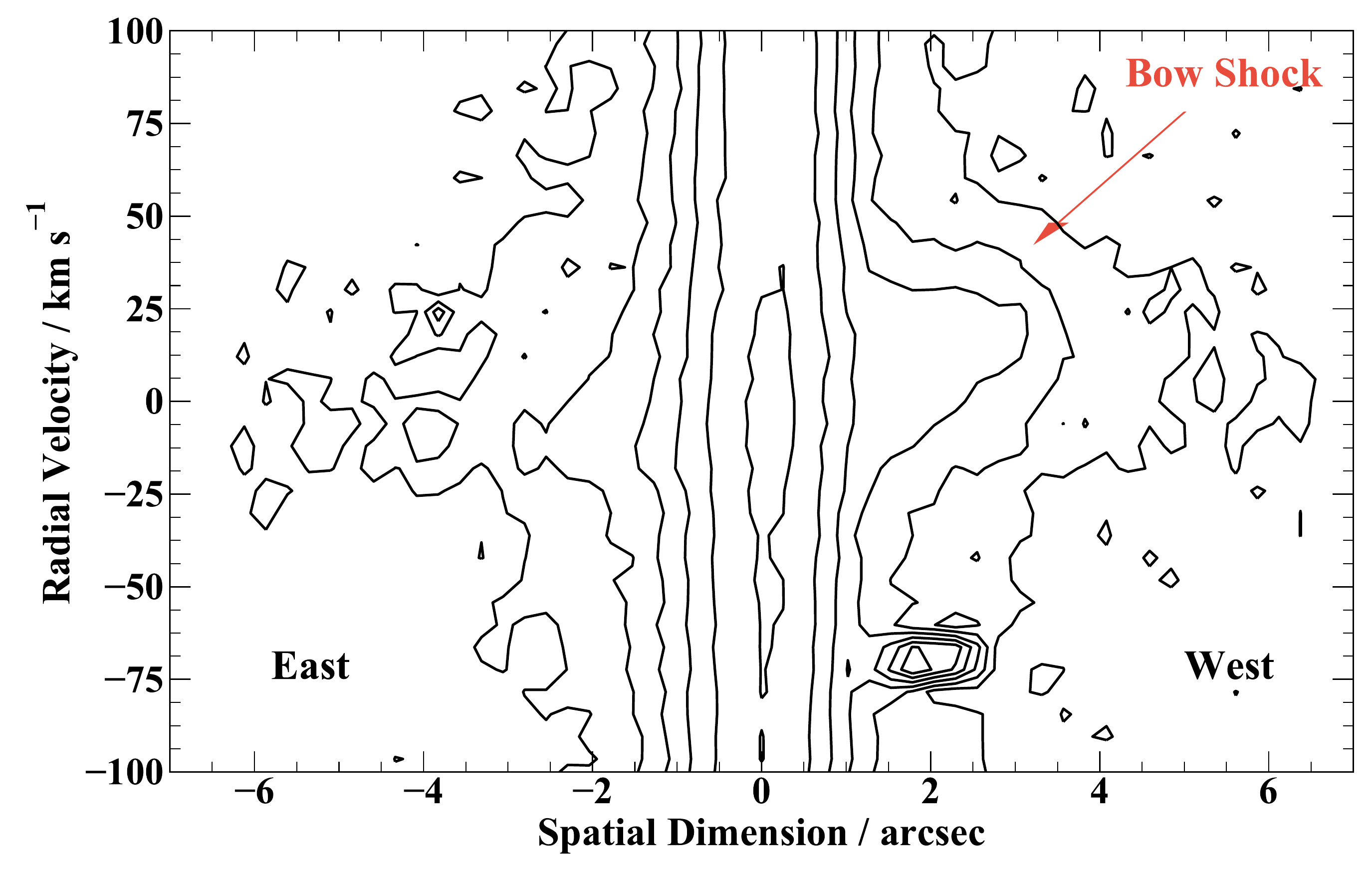}
\caption{
Deep H$\alpha$ imaging of \pnv. \textit{Top:} The centroid position for the Palomar plates is shown as the  circle. The proper motion vector (scaled for clarity) is shown and coincides with the direction of the diffuse emission form the bow shock. The Echelle slit is marked as the dotted box. \textit{Middle:} Stars have been subtracted to show only the extended emission. \textit{Bottom:} 2D H$\alpha$ profile obtained with the Echelle spectrograph. The wing extending towards the west coincides with the bow shock observed at later times with photometric observations.}
\label{fig:bow_shock}
\end{figure}

The extended emission depicted in the upper and middle panels of Fig.~\ref{fig:bow_shock} has a (Balmer) bow shock--like shape, formed by  our high space velocity object, which we believe is moving supersonically through the interstellar medium \citep{wil96}. Its shape and duration is very different from the recently discovered nova shells around CVs \citep{sea15}. In fact, the proper motion vector (displayed by the arrow in Fig.~\ref{fig:bow_shock}), coincides with the main axis of the bow shock. The head of the nebula is located about $\sim$7$^{\prime\prime}$ from the object and the size of the bow shock cone is $\approx$30$^{\prime\prime}$ along the sky plane with respect to the nebular symmetry axis. The cone open angle is about $\sim$60 degrees with its width and brightness quickly decreasing with increasing distance from the bow shock head. 

Similar bow shocks are frequently detected around OB--runaway stars \citep{vam88} and high--velocity pulsars \citep[and references therein]{yah17}. However, such bow shocks have also been observed in high-mass transfer CVs \citep[e.g. BZ Cam and V341 Arae][ respectively]{hea92,gea01,bond18}. It is generally accepted that the interstellar gas, compressed in bow shocks, is heated and ionised by the intense stellar radiation/wind and produces emission as a result of the collisional and/or the charge--exchange excitation of neutral hydrogen atoms in the post--shock flows, with a subsequent emission produced via bound--bound transitions \citep{cea80}. 

The time-scale associated with the presence of the bow shock suggests it was illuminated by the outburst and not material expelled in the outburst itself. Assuming very-fast outflows speeds for the ejecta ($v\sim0.1$c), the time-scale is of the order of $\sim80$ days to reach the observed distance ($~2\times10^{16}$ cm). This time-scale is larger than the limit imposed by the first detection of the bow shock ($<12$ days). Although high velocity outflows  can be achieved in energetic novae events \citep[e.g.][]{met14}, these velocities seem unlikely scenario for an outburst of short orbital period system \citep[which implies low-mass transfer rates][]{bre14}.

Furthermore, the high proper motion of the system (153~mas yr$^{-1}$) implies that, within an average recurrence time between outbursts (${\sim}30-50$ yr for WZ Sge-type CVs), the system would be displaced from the origin of a previous superoutburst by ${\sim}4.6$ arcsec. If this is the case, then the illumination of the apparent nebula would be centred on the location of the previous outburst, contrary to what is observed during the 2013 outburst. Moreover, the morphology of such expanding nebula would differ from a bow shock as shown in other outbursting systems like nova events \citep{sea15}. However, we do not observe any change in the illuminated nebula for over a month.

Therefore, in order to have a standing bow shock, the \pnv\ must have a quasi-continuous mass outflow. From momentum balance arguments \citep{wea77,vam88}, we can estimate a mass outflow rate of  $\sim10^{-11}$ M$_{\odot}$ yr$^{-1}$ to reproduce the observed bow shock \citep[assuming a ISM density of n$=0.2$ cm$^{-3}$, e.g.][]{hea92}. 
This value is roughly on the same order of magnitude as the putative mass transfer rate from the donor from evolutionary models \citep{kni11}, which suggests an outflow mechanism capable of expelling a significant percentage of the in-falling material with velocities of $\sim10^3$ \kms. Both requirements highly suggest that either a wind or jet-like outflow might be working in the system during the quiescent state.

\section{Spectroscopy}
\label{sec:spec}
We took a series of optical spectra during the evolution of the superoutburst of \pnv\@ from nearly maximum light down to a quiescent level (see Table~\ref{tab:speclog}). The spectral evolution of the emission and absorption lines as well as the shape of the continuum is summarised in the next two subsections. We also show in Fig.~\ref{fig:spec_evolution} a comprehensive graph of all the flux calibrated spectra (excluding the Echelle spectra), which illustrate the overall behaviour during this complex event.

\subsection{Outburst}
\label{sec:out}

The first 14 outburst spectra obtained with the  Echelle spectrograph, taken only three days (day 444) after the first report of the eruption \citep{ita13}, covered an interval of 2.5 hr. These high--resolution spectra show very strong double--peaked H$\alpha$ emission and a mixed emission and absorption component at the H$\beta$ line. Both emission line components have a FWHM $\sim350$ \kms\@ and are superimposed on a very broad absorption component ($\pm$ 1000 \kms). The broad component in H$\alpha$ is much weaker than in H$\beta$. No H$\gamma$ is present at all. Other features present on the spectra (not shown here) are weak \ion{He}{i} 5876 \AA\ and \ion{He}{i} 4471 \AA\ lines with mixed emission and absorption components. The double--peaked and low--velocity separation (peak to peak) of the H$\alpha$ line suggests an origin on the outer edges of the disc. 

\begin{figure*}
\includegraphics[angle=0,trim=0 0 0 0,width=17cm]{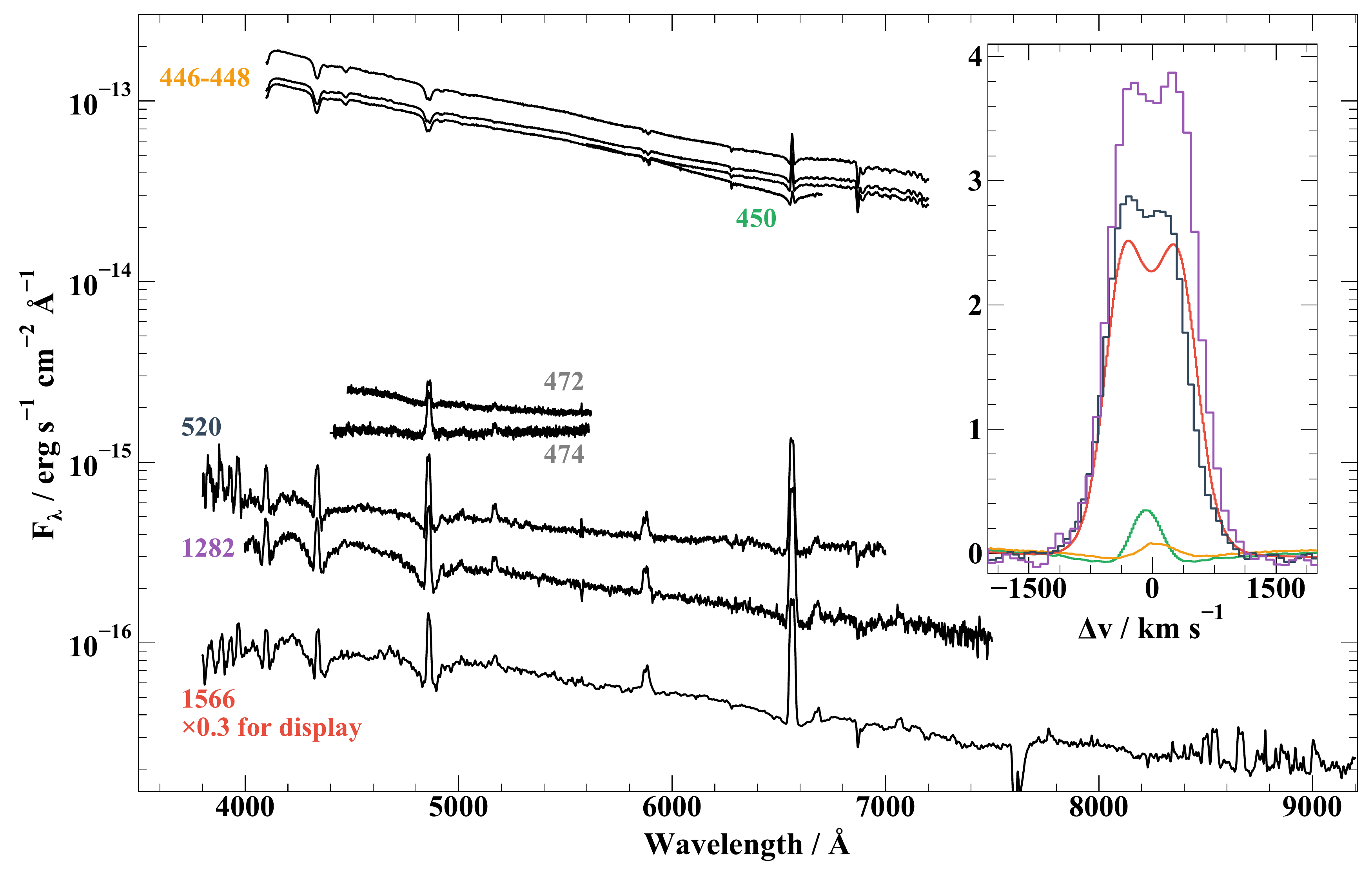}
\caption{
Optical spectral evolution throughout the superoutburst. We show the average spectrum for the corresponding epochs, labelled by our date notation, defined in Section~\ref{sec:neb}. The GTC spectrum has been smoothed for clarity. \textit{Inset:} The line profile evolution of H$\alpha$.}
\label{fig:spec_evolution}
\end{figure*}

On days 446--448 the B\&Ch spectra show a substantial change, as shown in the co--added spectra for the three nights in Fig.~\ref{fig:spec_evolution}. The broad absorption components dominate the Balmer series as well as \ion{He}{i} lines, except H$\alpha$ which show a single peak weak emission.  We measured the radial velocity shifts of the centre of this emission line (including night 450) using a single Gaussian with a FWHM$\sim6$ \AA. The first three nights show clear Doppler variations and a drift of the systemic velocity (from $\left\langle\gamma\right\rangle\sim+75$ to $-10$ \kms). However, our orbital coverage during these nights are very poor and no coherent modulation is observed. On day 450, the radial velocity measurements show a clear modulation with an apparent period coinciding with the early super--hump/orbital period. Due the short time--span of the data, we were not able to find a period value with enough reliability.

The inversion of Balmer emission lines to absorption is common for dwarf novae in outburst \citep[e.g.][]{clarke84,neu17}. This contrasts to what was observed during the superoutburst of the bounce--back candidate V455 And \citep{tea11}. In the latter, the Balmer lines, after an initial switch from emission to broad absorption, they suddenly reversed their course and shoot up back into emission. From the analysis of the width of lines and their radial velocities, \citet{tea11} concluded that this is evidence of evaporation and disc wind. This is noteworthy, because the observed bow shock of V1838~Aql supposes some kind of outflow from the object (see discussion in Section~\ref{sec:bowshock}). However, no similar wind or outflow features are seen at the beginning of the superoutburst in this case.

As the outburst progressed, the broad absorption line components shrink and eventually disappear. On the contrary, the H$\alpha$ line in emission broadens throughout the decline to its quiescent value of FWHM$\sim1200$ \kms\@ (see Section~\ref{sec:quiescence}). This is explicitly shown in the inset of Fig.~\ref{fig:spec_evolution}. On days 472--474 additional B\&Ch spectra were obtained, H$\alpha$ and H$\beta$ became again strong with double peak emission. 

\subsection{Quiescence}
\label{sec:quiescence}

\begin{figure}
\includegraphics[angle=0,trim=0.5cm .1cm 0.4cm 0.0cm,clip,width=\columnwidth]{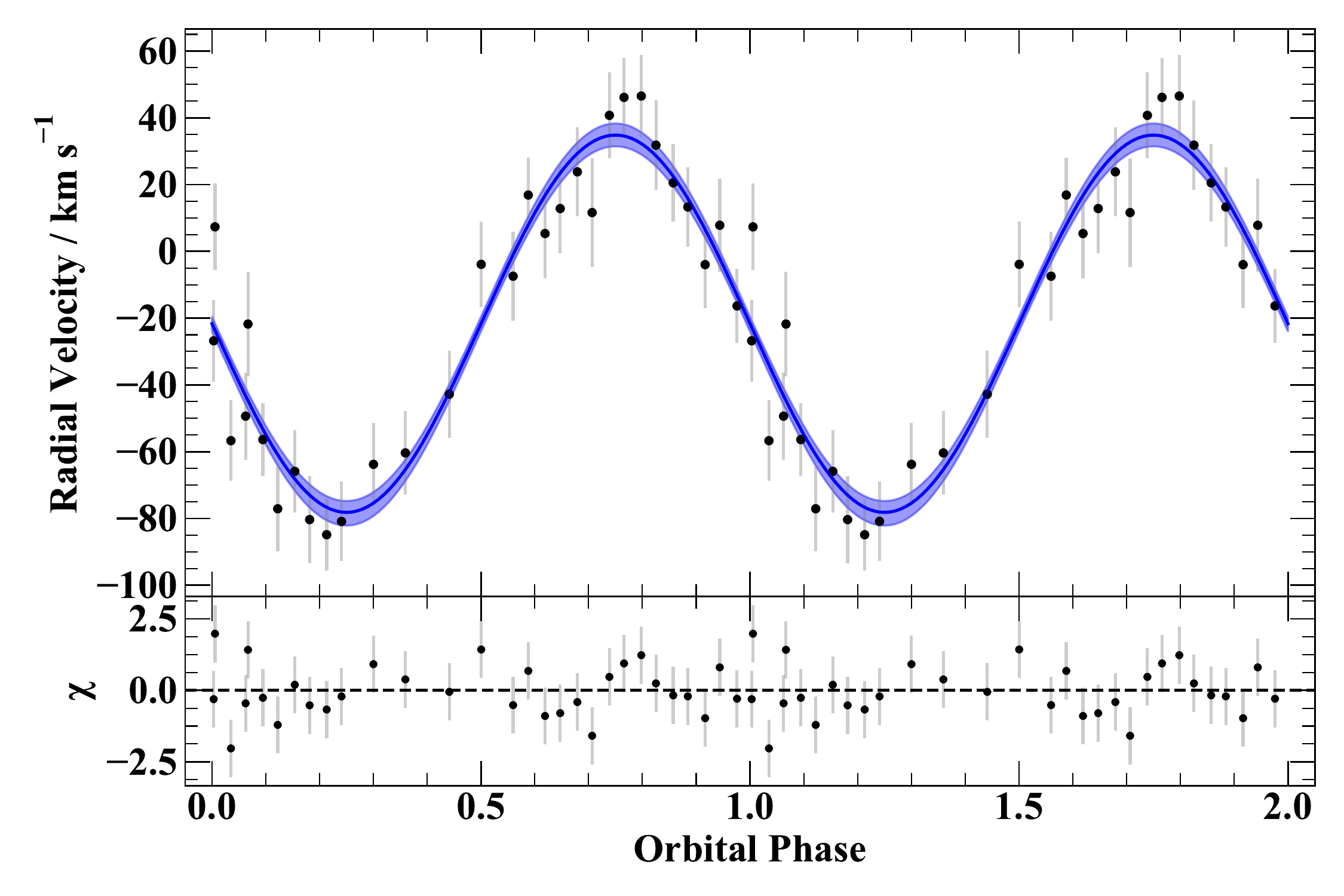}
\caption{
Radial Velocity curve of H$\alpha$ in quiescence for the GTC spectra. Radial velocities were obtained via 2--Gaussian technique. The best fit is shown as the blue line and 1$\sigma$ error bars have been scaled so $\chi^2_{\nu}=1$.}
\label{fig:radial_curve}
\end{figure}
The system returned to its quiescence state after 500 days, thus presenting an opportunity to analyse the system in its quiescent state. The full optical spectrum, taken over 1000 days after the outburst, shows a clear presence of accretion disc emission lines superimposed on the WD broad absorption lines, as shown in Fig.~\ref{fig:spec_evolution}. Time-resolved spectroscopy however, was focused on the region around the H$\alpha$ line, where the least contribution of the WD broad absorption lines is observed. Hence, we were able to apply a two Gaussian method \citep{sha86,horne86b} to determine the radial velocity of the primary. We made an interactive search for the optimal width and separation of the Gaussians in a grid between a FHWM of 500--1000 \kms ~in 50 \kms ~steps and 500--2600 \kms ~in 100 \kms ~steps, respectively \citep[for further explanation of this method see][]{her17}. At every combination, we performed a fit of the radial velocities, $V(t)$, to a circular orbit:

\begin{equation}
V(t) = K_{em}\sin\left[2\pi(\phi-\phi_0)\right] + \gamma\,,
\label{eq2}
\end{equation}
where $K_{em}$ is the semi--amplitude, $\phi_0$ is the phase offset between the spectroscopy and photometric ephemeris, and $\gamma$ is the systemic velocity.  
We used the minimum in the diagnostic quantity $\sigma_K{_{em}}/K_{em}$, where $\sigma_K{_{em}}$ is the 1$\sigma$ uncertainty on the semi--amplitude, to determine the optimal solution \citep{sha86}. 

Initially, we have used an arbitrary zero-point close to the observations to calculate the phase offset and have assumed the orbital period of the early superhumps. We obtain an orbital solution with $P_{\rm{orb}}=0.0545\pm0.0026$ days, K$_{em}=53\pm3$ \kms\ and $\gamma=-21\pm3$ \kms\ as shown in Fig.~\ref{fig:radial_curve}. This spectroscopic orbital period is consistent with the early superhump period found during the superoutburst  \citep[$P=0.05698 \pm 0.00009$ d,][]{kato14a,ech18}. 

After correcting for the phase offset to fix the zero point of the inferior conjunction of the secondary, we can construct the spectroscopic ephemeris:

\begin{equation}
T(HJD) = 2457566.5290(5) + 0.0545(26)E.
\label{eq3}
\end{equation}

\subsection{Doppler tomography}
\label{subsec:doptom}

\begin{figure*}
\includegraphics[trim=2.5cm 7.5cm 2.0cm 4.5cm,clip,width=18cm]{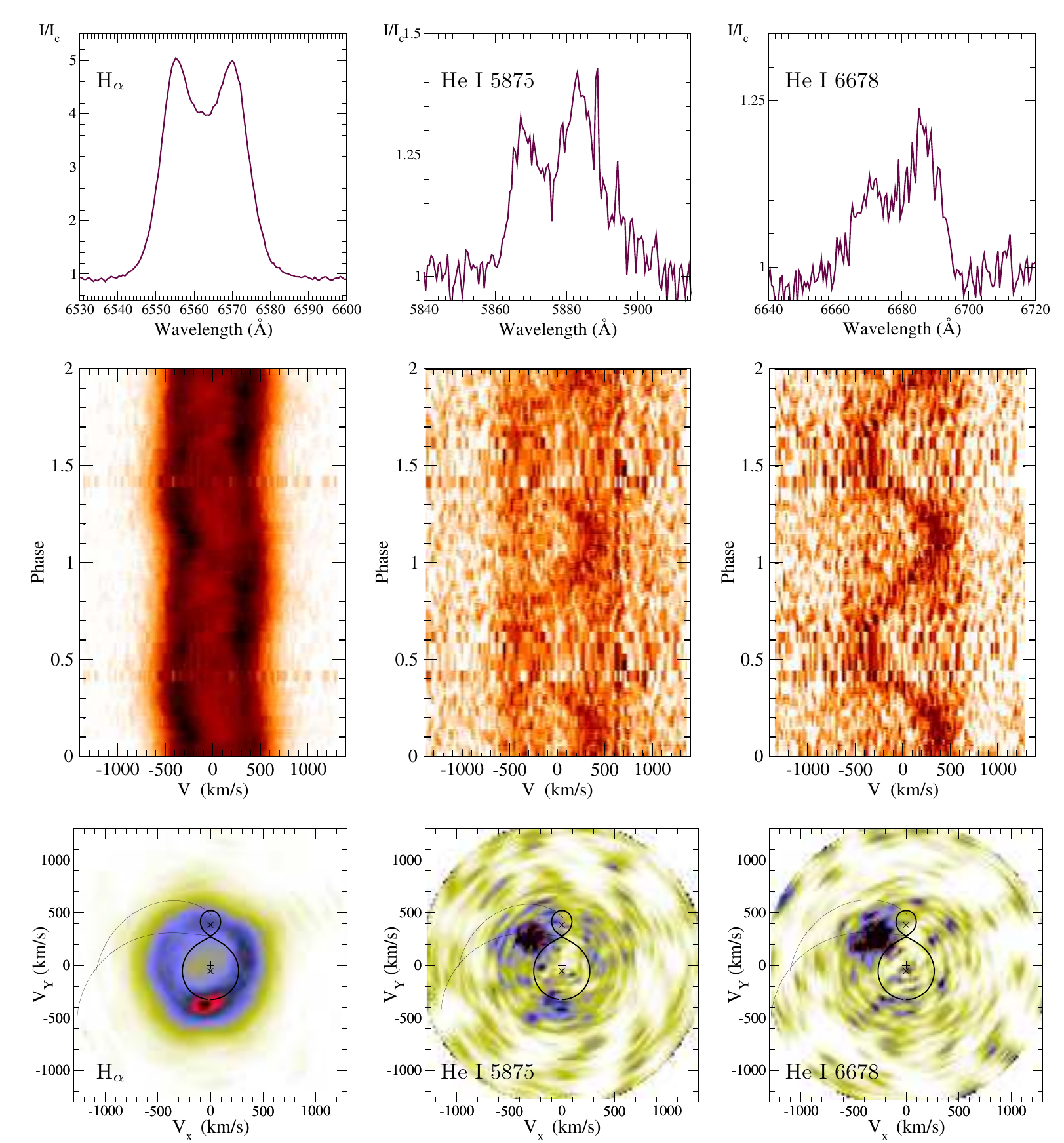}
\caption{
Time-resolved spectra of \pnv\@ during quiescence for H$\alpha$, He~I 5875\AA\@ and 6678\AA. For each line we show the median line profile (\textit{top}), trail spectra (\textit{middle}), and Doppler tomogram reconstructed for each line (\textit{bottom}). The Roche lobe surface for the donor (solid line) and primary (dotted line) were calculated using following orbital parameters: $i = 60^{\circ}$, $q=0.1$ and $\gamma = -28$ \kms. The Keplerian and ballistic trajectories in the figure are marked as the upper and lower curves, respectively. The crosses are the velocity (from top to bottom) of the secondary star, the centre of mass and the primary star.}
\label{fig:doppler}
\end{figure*}

The quiescent spectrum of \pnv\ contains emission lines that allow us to study the structure of the accretion flow in H$\alpha$ and He I 5875~\AA\@ and 6678~\AA, as shown in Fig.~\ref{fig:doppler}.  We have used Tom Marsh's \caps{molly} software package\footnote{\url{http://www2.warwick.ac.uk/fac/sci/physics/research/astro/people/marsh/software/}} to normalise and subtract the continuum of each individual spectrum, as well as re--bin in equal velocity bins. All three lines show a distinct double--peak profile which suggests the presence of an accretion disc, shown in the mean profiles in the top panel of Fig.~\ref{fig:doppler}. In addition, the lines contain an extra component with a semi--amplitude of $\sim200$ \kms, which modifies the symmetry of the median profile (most evident in the \ion{He}{} lines and are clearly seen in the trail spectra (mid--panels, Fig.~\ref{fig:doppler}). However, we note that the  extra component in H$\alpha$ seems to be shifted in orbital phase with respect to both \ion{He}{i} lines. 

The time--resolved spectroscopic data allow us to study the structure of the accretion disc via Doppler tomography \citep{horne86}. We have used the ephemeris obtained via the H$\alpha$ wings (see Section \ref{sec:quiescence}) to calculate the tomograms for each individual emission line. We employed the Doppler tomography package \caps{trm-doppler}\footnote{Available at \url{https://github.com/trmrsh/trm-doppler}} to produce the maps shown in the bottom panels of Fig.~\ref{fig:doppler}. The H$\alpha$ tomogram reveals a clear accretion disc as well as an additional component at $V_x,V_y \approx [0,-500]$ \kms. Surprisingly, we find a lack of distinct emission of the hot spot in H$\alpha$, commonly observed in most CVs. The \ion{He}{i} tomograms produce a sinusoidal contribution superimposed on the weaker accretion disc. The location of the emission in \ion{He}{i} coincides with the expected position in velocity space of the hot spot, where the ballistic trajectory of material ejected from the L1 point intersects the outer edge of the accretion disc.

The fact that the ephemeris is obtained via the wings of the H$\alpha$ line and simultaneously providing a consistent position for He I hot spots, indicates that the H$\alpha$ position might be real and not an error of the zero--point used. However, we need more radial velocity observations of \pnv\ in quiescence to lock down the real zero point.

\subsection{White dwarf temperature}
\label{sec:wd}
\begin{figure}
\includegraphics[angle=0,trim=0.0cm .6cm 0.0cm 0.0cm,clip,width=\columnwidth]{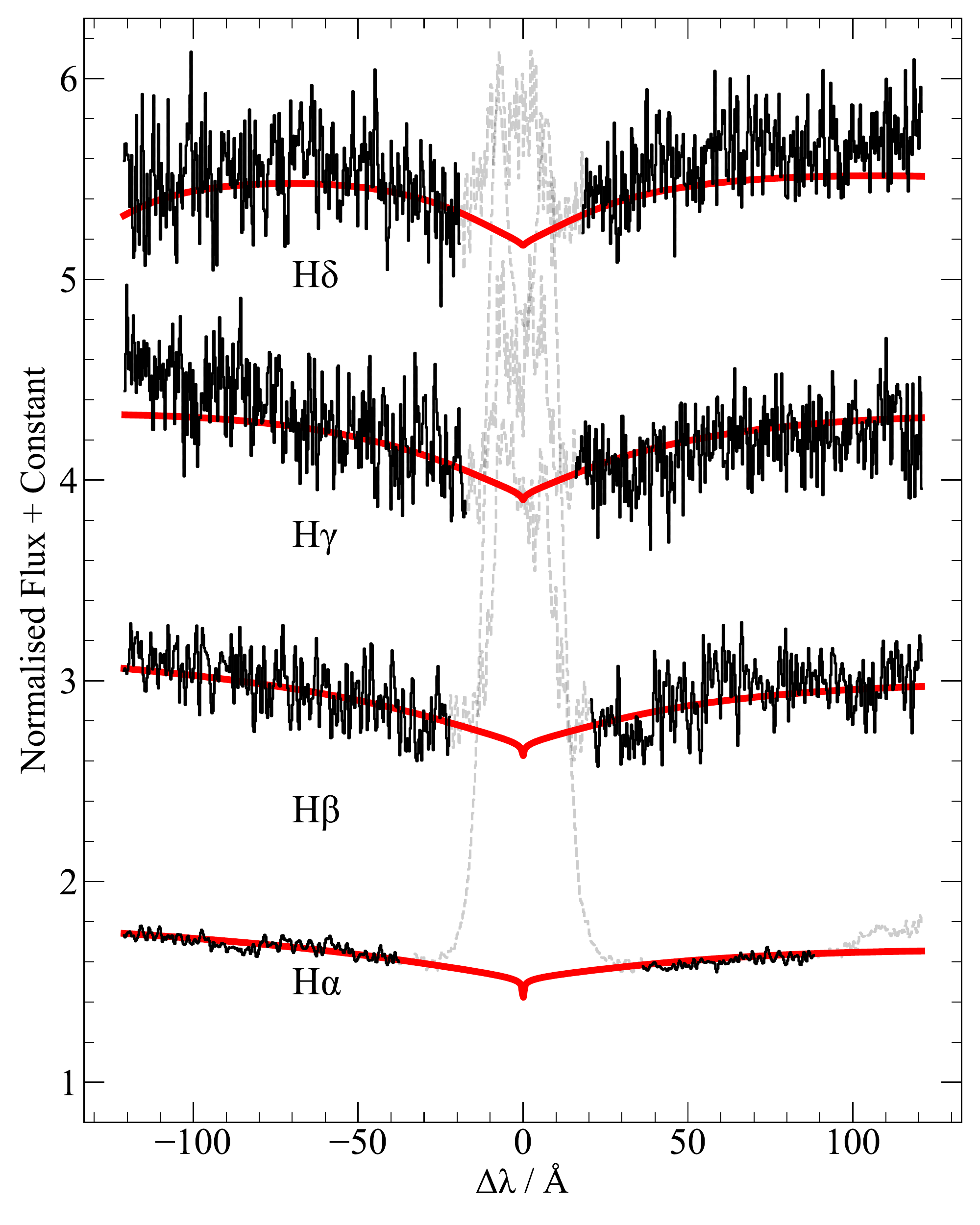}
\caption{WD atmosphere model best fit (red) of the Balmer series wings. The emission core originating from the accretion disc has been masked. An arbitrary offset has been applied for clarity.}
\label{fig:wd_temp}
\end{figure}

In quiescence, the optical spectrum of \pnv\@ is dominated by the broad absorption hydrogen lines arising from the atmosphere of the WD. However, even at these low luminosities, the accretion disc can contribute a significant fraction of the continuum and line emission at optical wavelengths \citep[e.g.][]{avi10,her16}. Thus, in order to retrieve the physical parameters of the WD, we have modelled the SED in quiescence, $F_{\textrm{q}}$, as a combination of a WD atmosphere model $F_{wd}$ broadened to the instrumental resolution ($R\sim2500$) and a power-law component:
\begin{equation}
F_{\textrm{q}}(\lambda) = \frac{R_{wd}}{d}^{2}F_{wd} + A\cdot \left(\lambda\right)^{\Gamma},
\label{eq:wd_fit}
\end{equation}
where $R_{wd}$ is the radius of the WD, $d$ the distance, $A$ is the normalisation factor of the power law and $\Gamma$ is the power law index. We have masked the cores of the Balmer lines and fitted the range between 3900 -- 7500 \AA. We fixed the distance from the \textit{Gaia} DR2 estimate, $R_{WD} = 0.01$ R$_{\odot}$ (radius for a 0.8 \msun\ and $\log(g)=8.35$ WD star) and left $\Gamma$ and the normalisation as free parameters. We then performed a grid search over a set of WD atmosphere models made with \caps{tlusty/synspec} \citep{hub95,hub17}. This grid consisted of a single value of surface gravity $\log(g)=8.35$ which corresponds to the mean WD mass for CVs \citep{zor11} and a range of effective temperatures, T$_{eff}$, 8000--30000 in steps of 100 K.  The best fit model is shown in Fig.~\ref{fig:wd_temp}, where we show the broad absorption wings of the Balmer series. We used this relation and obtained T$_{eff}=11,600\pm400$ K for the WD temperature and a power law index of $\Gamma=-1.40\pm0.1$. The accretion disc contributes 41$\pm5$\% of the optical light in this region, similar to other short orbital period systems \citep{avi10,zhar13}. The contribution of the disc is lower at later epochs ($\sim10$\%), when the system reaches its pre-outburst flux level, assuming the WD temperature does not cool down significantly during this period. However, the lack of spectroscopy at later date prevents us to confirm this scenario.

We can also approximate the observed quiescent accretion disc spectrum using a spectrum of the optically thin slab which mimics the radiation of the quiescent disc.  We used simplified one-zone approximation for the spectrum calculation, considering homogeneous slab. The spectrum was computed using the well known solution for the homogeneous slab $F_{\lambda} = \pi B_{\Lambda} (1- \mathrm{exp}(-\kappa_{\lambda}(T,\rho)*\rho*z))$, where $z$ is the geometrical slab thickness, $\rho$ is the matter density in the slab, and $T$ is the slab temperature. The true opacity $\kappa_\lambda$ was computed in LTE approximations using corresponding subroutines from Kurucz's code \caps{atlas} \citep{Kurucz:70,Kurucz:93} adopted by V.~Suleimanov \citep{Suleimanov:92,IS:2003,SW:07}. The true opacity means that we considered only bound-bound, bound-free, and free-free transitions and ignored electron scattering. The slab parameters were tuned by hand to find the ones for which the model shows the best agreement with the observations. The accepted parameters are $T = 28\,4000$\,K, $z=5\times 10^6$\,cm, $\rho =5.7 \times 10^{-10} $ g cm$^{-3}$, and $V \sin i = 518$\,km\,s$^{-1}$. The derived size of the slab is $R_{\rm sl} \sqrt{\cos i} = 2.34 \times 10^9$\,cm. We assumed that the slab has the solar chemical composition. We note that the calculated slab spectrum is in agreement with the power law fit obtained above, but shows less contribution to the total system flux in the NIR wavelengths.

Initially, we independently calculated the distance to \pnv\ previous to the \textit{Gaia} DR2 release. Using the~mean value of the absolute magnitude of 340 White Dwarfs in binary systems from the catalogue of McCook--Sion \citep{mccook99}, we obtained a mean equal to $\langle M\rangle =12.61$. We assumed that the brightness of the accretion disc of a period bouncer system contributes from $\sim20-40$\% of the brightness of the primary star \citep{avi10,zhar13}. The faintest magnitude observed for \pnv\ is ($V=18.6$), combined with the mean absolute magnitude obtained from the McCook--Sion catalogue, we find a distance interval $187 \pm 50$ pc, consistent with the distance measured afterwards by \textit{Gaia}. 

The temperature of the WD is both consistent with theoretical predictions \citep{town03,kni11} and observational estimates for systems close to the period minimum \citep{zat15, pala17}. Furthermore, the measured $T_{eff}$ lies within the instability strip, where pulsations are observed in isolated \citep{gia06} and accreting WDs \citep[e.g. GW Lib,][]{szkody10}. Future high-time resolution photometry in the ultraviolet or blue optical bands might provide a new candidate to study non-radial pulsations in WDs \citep[e.g.][]{uthas12}.
 
\section{Discussion}

\subsection{Approaching or receding the period minimum?}
Both theoretical predictions and observations of the CV population show a sharp cut--off in the orbital period distribution at about 80 min,  the so-called period minimum \citep{rak98,gan09}.
\pnv, with an orbital period about ${\sim}82$ min \citep{kato14a,ech18}, is located within the spread of systems around the period minimum and therefore makes it difficult to establish if the system is approaching or leaving it (i.e. period bouncer). 

The light curve of its first (and only) recorded superoutburst (with no normal outbursts detected) can help address this question. Its morphology (e.g. amplitude, duration), has similar features observed in other WZ~Sge--type systems. We observe a sudden drop in flux (both in optical and X-rays, see Fig.~\ref{fig:aavso-figs}), followed by a gradual decrease back to its quiescent level \citep[likely from the steady cooling of the WD, e.g.][]{neu17}. In particular, we note the lack of rebrightenings in \pnv\ which are present in most WZ~Sge--type stars.

Following the empirical morphological classification of superoutburst light curves proposed by \citet{iea06}
, \pnv\  belongs to the {\it type} D morphology (i.e. no rebrightenings and sudden flux drop). 
\citet{kat15} ascribes this morphology classification to an evolutionary sequence from pre- to post- period minimum systems (C:D:A:B:E) and points out that {\it type} D might be closely associated with systems around the period minimum, but still in the upper branch of the $q-P_{orb}$ branch. This is consistent with the larger contribution in the NIR by the donor (see Section~\ref{sec:donor}), which suggests \pnv\ to be a pre-bounce system.

Another way to discern if a system is a period bouncer is to look for distinct features in quiescence such as permanent double-hump light curve as well as a spiral arm structure in their Doppler tomography \citep{zat15}.
\citet{ech18} presented two light curves of \pnv\ obtained in 2018 which do not show a double-hump modulation. However, the lack of a precise ephemeris prevents to further explore the presence of more subtle modulations. With regards to a spiral pattern, we observed a distinct feature in the accretion disc in a peculiar position of the velocity space (see H$\alpha$ tomogram in Section~\ref{subsec:doptom}). This feature however, is different to the ones observed in good period bouncer candidates as V406 Vir \citep{zhar08} or EZ Lyn \citep{zhar13}, where a dual-emitting component was associated to spiral patterns in the disc. On the other hand, it resembles more to those found in HT Cas \citep[e.g.][]{nzb16}, as discussed further in Section~\ref{subsec:doptom}. In any case, until the real phasing of the system is found no definitive conclusion on the origin of these structure can be drawn.

Period bouncers have small mass ratios ($q\lesssim0.1$). In consequence, the more massive WD should present small semi-amplitudes, $K_1$, in their radial velocity curves. In order to test this for \pnv\, it is necessary to estimate the inclination angle of the system. 
The quiescent spectrum clearly shows a double--peak emission on the Balmer series as well in the \ion{He}{i} which is observed in accretion discs with inclination angles $i\gtrsim15^{\circ}$ \citep{horne86}. Also, the lack of eclipses in the photometry imposes an upper limit, for reasonable mass ratios of period bouncers, $i\lesssim81^{\circ}$ \citep[for $q\simeq0.05$,][]{bail90}. 
In addition, \pnv\ shows small peak-to-peak separations (${\sim}650-700$ km s$^{-1}$ for H$\alpha$) in contrast to eclipsing systems at similar orbital periods \citep[e.g. ${\sim}1300$ \kms\ for SDSS~J1433+0038,][]{tulloch09}. Comparing this with other eclipsing short--periods CVs, we conclude that the inclination angle is relatively small ($i{\sim}30^{\circ}$)
Thus, the {\it real} not projected semi-amplitude is very high ($K_1{\sim}100$~\kms). Assuming the $q$ determination from superhumps  \citep{ech18}, the donor in \pnv\ would lie above the sub-stellar threshold. Again, this would argue for a pre-bounce system.



\subsection{Low--mass star or sub--stellar donor?}
\label{sec:donor}

\begin{figure*}
\includegraphics[angle=0,trim=0.0cm 0.5cm 0.0cm 0.0cm,clip,width=18cm]{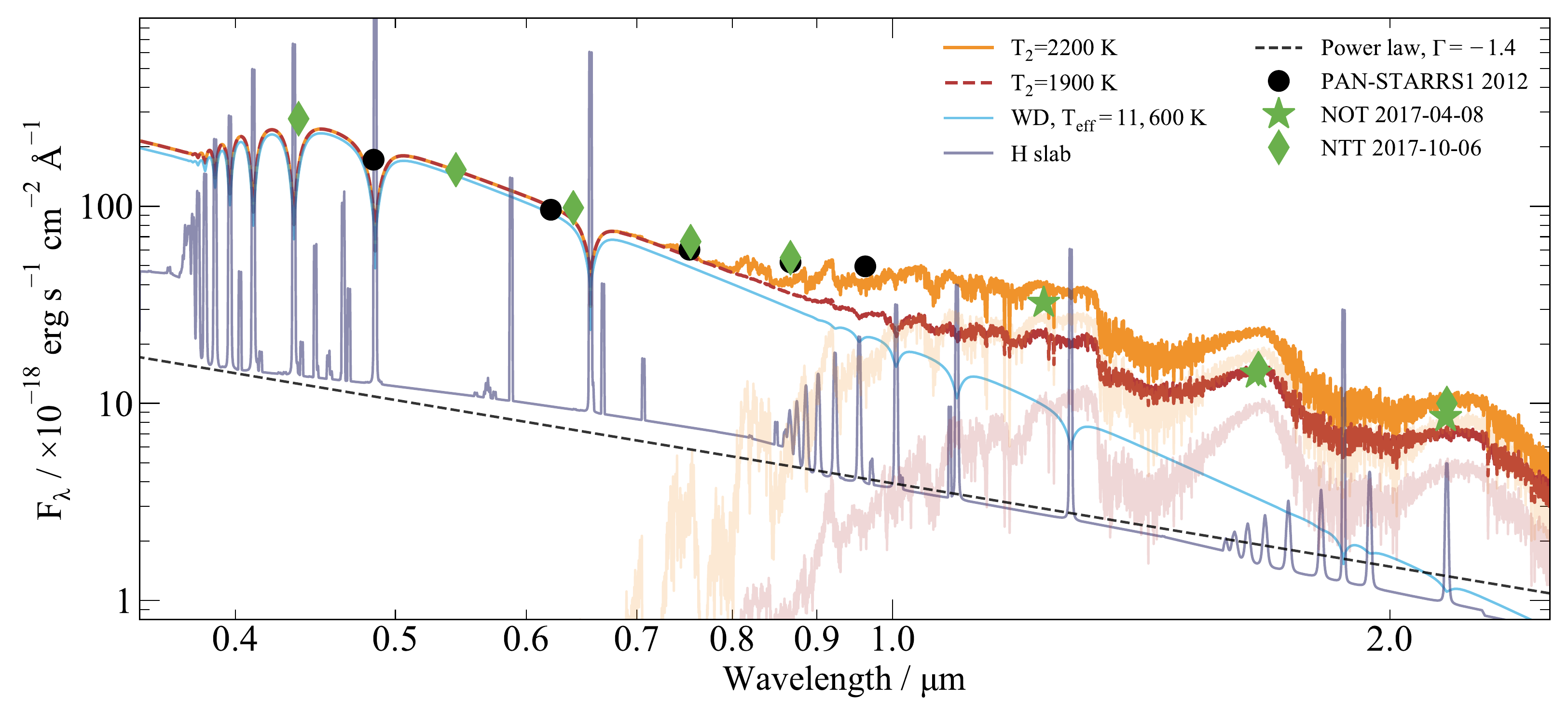}
\caption{Broadband SED of \pnv\ in quiescence. We show the multi-band photometry before (PAN-STARRS) and post-outburst (NOT and NTT). Two models have been scaled to compare the NIR contribution of the donor using low-mass star atmospheres of 2200 K (red) and 1900 K (orange), corresponding to a pre and post-bounce system \citep{kni11} respectively. The WD and power law index are taken from the Balmer wing fit. The power law contribution has been scaled by hand to match the quiescent level ($\sim10$\% contribution). The thin slab model is also scaled down and show as reference.}
\label{fig:broad_sed}
\end{figure*}

The large wavelength range of the GTC spectrum allowed us to search for evidence of the donor (e.g. absorption features) as shown in Fig.~\ref{fig:broad_sed}. In particular, we searched in the region between 0.7--1 $\mu$m where the SED of the donor should start to contribute a significant percentage of the system's light. However, we find a wide range of the spectrum to be mostly dominated by broad emission Paschen lines. In the non--contaminated regions, we find no absorption features associated with the donor.

Despite the lack of evidence of features of the donor star in the optical region, the donor continuum should contribute as significant fraction at longer wavelengths. We used the multi-coloured broadband photometry in a true quiescent state to illustrate the type of donor expected from its NIR properties\footnote{We note that at the time of the GTC spectrum (day 1566), the system is slightly brighter than its pre-outburst state taken by PAN-STARRS1 in 2012 and a later observation taken with the NTT (day 2032). Unfortunately, we do not have information of the NIR flux close to the GTC epoch. This suggests that even 3 years after the superoutburst, either the WD is still cooling or the disc remains brighter.}. Theoretically, the orbital period suggests that the donor is likely to have a spectral type later than M6 \citep[T$_2\lesssim2\,400$ K,][]{kni06,kni11} depending whether the system lies before or after the period minimum. We show in Fig.~\ref{fig:broad_sed} two models using the fit done in the optical region (where the donor contribution is minimal) scaling down the power-law contribution to match the photometry and adding a low-mass star atmosphere model of $1\,900$~K and $2\,200$~K \citep{bar15} appropriate for a pre- and post-period minimum system at P$_{orb}{\sim}82$ min, respectively. No fit has been performed and only presented here as reference. 

The photometry shows a flux increase starting at $\sim$8000~\AA, deviating from just the simple addition of a WD model and the power law. This excess in addition to the better match of the NIR fluxes to the hotter component  suggests that \pnv\ is indeed approaching the period minimum. However, a consistent fit with quasi-simultaneous data is needed to deliver a more accurate estimate of the properties of the donor. Thus, \pnv\ is an ideal candidate for NIR time--resolved spectroscopy \citep[e.g.][]{her16}, which would render a fully independent measurement of the mass ratio and confirm or reject its sub--stellar nature. This is particularly important since few systems have been characterised close to the period minimum, where the transition from stellar to sub-stellar regime is expected.

\section{Summary}
\label{sec:conclusions}

We have presented a long--term photometric and spectroscopic study of the 2013 superoutburst of \pnv\ from its peak to quiescence. The morphology of the outburst in combination with previous photometric estimates \citep{kato14a,ech18} allow us to confirm \pnv\ as a short orbital period system and provide updated ephemeris.
A few days after the peak of the outburst, we discovered extended emission around the object, as observed in the high-resolution spectroscopy. Further deep H$\alpha$ revealed an illuminated bow shock consistent with the proper motion of the object ($123\pm 5$ \kms). Although the origin of the material that creates the bow-shock is unclear, we conclude that a quasi-continuous outflow of material (${\sim}1000$ \kms) is required to sustain a standing bow shock with the ISM.

In quiescence, we obtained time--resolved spectroscopy which allowed us to determine a semi--amplitude of the primary $K_1=53\pm3$ \kms. Doppler tomography in H$\alpha$ revealed an emission component inconsistent with the ballistic trajectory of the accretion stream observed in the \ion{He}{i} lines. Further observations and refinement of the ephemeris are needed to discern the origin of this emitting component.
The broad band spectroscopy, allowed us to infer the effective temperature of the primary $T_{eff}=11,600\pm400$ K. This is consistent with theoretical expectations \citep{town03} as well as observational constrains on similar systems \citep{pala17}. 

A discussion is made on the possible period-bouncer nature of the object. The broadband optical and NIR photometry suggests that \pnv\ is approaching the period minimum limit and hosts a $>2\,000$ K donor. However, we point out that simultaneous optical and infrared time-resolved spectroscopy (or photometry) needs to be performed, in order to measure the radial velocity curves of both components and to determine the spectral type of the donor star and its possible sub-stellar nature.

\section*{Acknowledgements}
The authors are indebted to DGAPA (Universidad Nacional Aut\'onoma de M\'exico) for financial support, PAPIIT projects IN111713, IN122409, IN100617, IN102517, IN102617, IN108316 and IN114917. JVHS is supported by a Vidi grant awarded to N. Degenaar by the Netherlands Organization for Scientific Research (NWO) and acknowledges travel support from DGAPA/UNAM. JE acknowledges support from a LKBF travel grant to visit the API at UvA. VN acknowledges the financial support from the visitor and mobility program of the Finnish Centre for Astronomy with ESO (FINCA), funded by the Academy of Finland grant No. 306531. GT acknowledges CONACyT grant 166376. E. de la F. wishes to thank CGCI--UdeG staff for mobility support. VS thanks Deutsche Forschungsgemeinschaft (DFG) for financial support (grant WE 1312/51-1). His work was also funded by the subsidy allocated to Kazan Federal University for the state assignment in the sphere of scientific activities (3.9780.2017/8.9).

We thank Tom Marsh for the use of \caps{molly}. We acknowledge with thanks the variable star observations from the {\it AAVSO International Database} contributed by observers worldwide and used in this research. We acknowledge the use of public data from the Swift data archive. This research made use of \caps{astropy}, a community--developed core \caps{python} package for Astronomy \citep{Astropy-Collaboration:2013aa}, \caps{matplotlib} \citep{Hunter:2007aa} and \caps{aplpy} \citep{Robitaille:2012aa}. Based (partly) on observations made with the Gran Telescopio Canarias (GTC), installed in the Spanish Observatorio del Roque de los Muchachos of the Instituto de Astrof\'isica de Canarias, in the island of La Palma (GTC7-16AMEX). Partly based on observations made with the Nordic Optical Telescope, operated by the Nordic Optical Telescope Scientific Association at the Observatorio del Roque de los Muchachos, La Palma, Spain, of the Instituto de Astrofisica de Canarias. The data presented here were obtained in part with ALFOSC, which is provided by the Instituto de Astrofisica de Andalucia (IAA) under a joint agreement with the University of Copenhagen and NOTSA. The results presented in this paper are based on observations collected at the European Southern Observatory under programme ID 0100.D-0932. We thank the day and night--time support staff at the OAN--SPM for facilitating and helping obtain our observations. This work has made use of data from the European Space Agency (ESA) mission {\it Gaia} (\url{https://www.cosmos.esa.int/gaia}), processed by the {\it Gaia} Data Processing and Analysis Consortium (DPAC, \url{https://www.cosmos.esa.int/web/gaia/dpac/consortium}). Funding for the DPAC has been provided by national institutions, in particular the institutions participating in the {\it Gaia} Multilateral Agreement. We thank J. van den Eijnden for help on \textit{Swift's} DDT proposal.


\appendix
\section{\textit{Gaia} Posterior distributions}
\label{sec:A1}

\begin{figure*}
	\includegraphics[width=10cm]{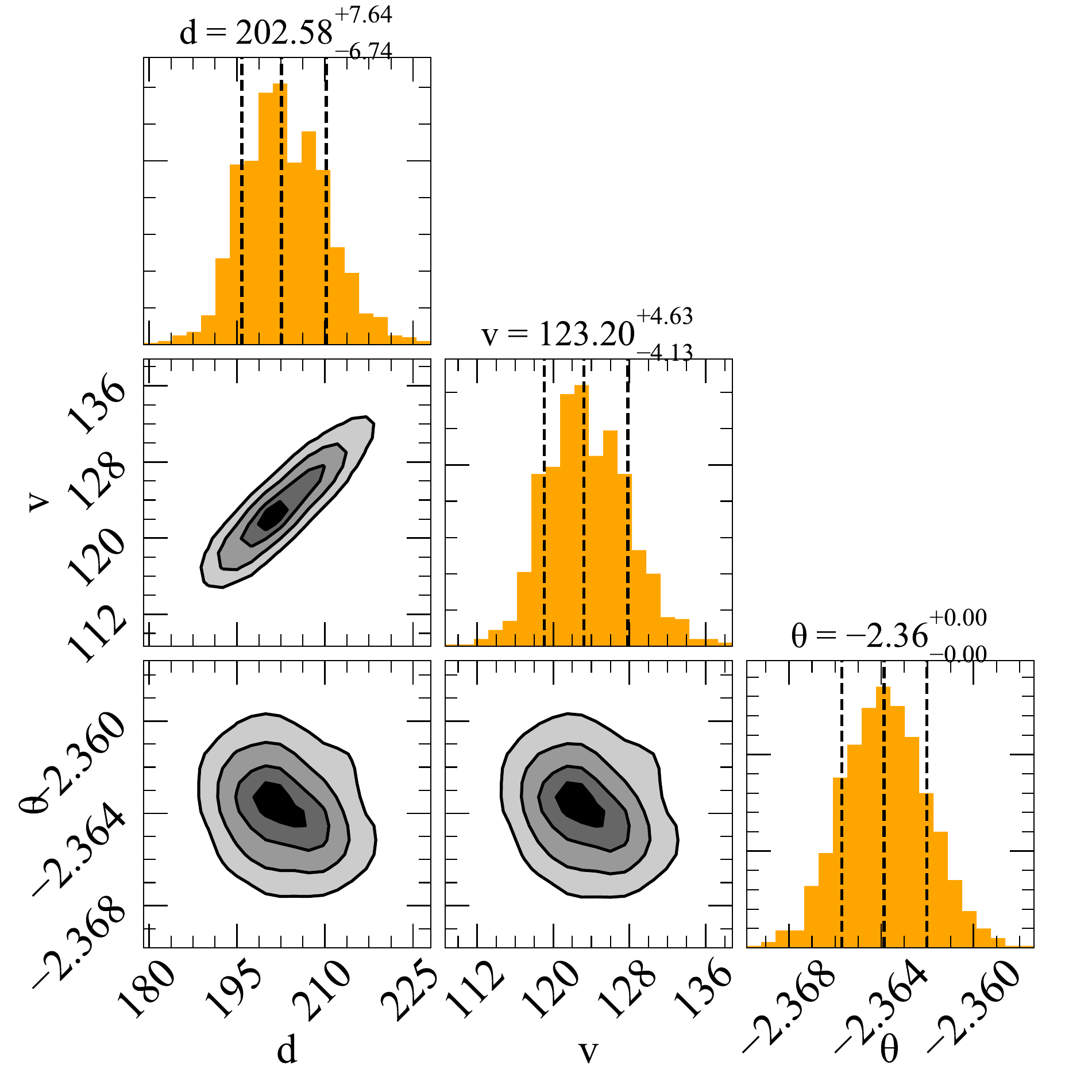}
    \caption{Posterior probability distributions for the \textit{Gaia} distance and tangential velocity. Colour scale contours show the joint probability for every combination of parameters. Units for each parameter are: distance in pc, velocity in \kms\ and angle in radians. Contours represent the 0.5$\sigma$, 1$\sigma$ , 2$\sigma$ and 3$\sigma$ levels. Marginal posterior distributions are shown as histograms with the median and 1$\sigma$ marked as dashed lines. }
    \label{fig:posteriors}
\end{figure*}

We present the joint and marginal posterior distributions of the \textit{Gaia} distance and tangential velocity, shown in Fig.~\ref{fig:posteriors} (see Sec.~\ref{sec:bowshock}). We used \caps{corner.py} \citep{corner} to visualise the MCMC chains.

\bsp

\label{lastpage}

\end{document}